# To the Fluid Motion Dynamics

S. L. Arsenjev[1]

*Physical-Technical Group*
*Dobrolubova Street, 2, 29, Pavlograd, Dnepropetrovsk region, 51400 Ukraine*

A fluid motion through the flow element is presented in the kind of an auto-oscillating system with the distributed parameters: mass, elasticity, viscosity. The system contains a self-excited oscillator and possesses a self-regulation on base of the intrinsic voluntary negative servo output feedback. The interaction dynamics of the submerged jet, out-flowing out of a flow element, with the homogeneous for it surrounding medium as well as dynamics of evolution of the spiral-vortex structures, appearing at the relative motion of fluids, are described for the first time. The conceptual model of a flow in the kind of the completed system of the cause and effect relationship, that presents the fluid motion process as a result in interaction of the motive power, applied from without, and the internal wave mechanism, self-regulating the flow structure and its intensity, is created.

**Introduction**
In our previous articles [1,2] the authors have presented the computation results, that physically adequately and mathematically correctly for the first time reproduce the flow structures, the fluid state and motion parameters as well as the weight-flow characteristics of the flow elements in the kind of a pipe, nozzle. The obtaining of such qualitatively new information on the base of analytical description of a gas stream dynamics, that is free from a mathematical scholasticism and an experimental empiricism, became feasible owing to using of the Physical Ensemble method (below PE-method). In the simplest applied sense the PE-method envisages the creation of the special conceptual structures accordingly to the number of possible degrees of a motion freedom of the considered object (in the given case – a gas), and then conjugation of the special conceptual structures in the integral consistent and indiscrepant – logically and physically – structure with taking into account of properties of the object (gas), the counterface (the flow element) and the motive power (pressure drop), applied from without to the flow system. Further PE-method envisages a possibility to translate the verbal form of the integral conceptual structure into a form of the solvable analytical algorithm and then to work out a program-solver on this base. Such program is, in particular, VeriGas- program in 01.01…01.04 versions, created by the authors by the middle of 2001.
Combination of such properties as a mass density (inertness), reversible volume compressibility (volume elasticity) and an internal friction allows supposing the real fluid motion as an oscillatory process.
In one's turn, a flow property to reproduce the flow structures, the fluid state and motion parameters and the weight-flow characteristics of the flow element allows to suppose the real fluid motion as the self-regulated process.
Some of the above mentioned features of the fluid motion were took into account in VeriGas program-solver by means of PE-method and allowed by means of the computing experiment to reproduce such details of the gas stream motion that either were not detected in the special laboratory experiments or were not explained physically adequately.

---

[1] Phone: (+38 05632) 40596
E-mail: usp777@ukr.net

**Approach**

Some principles of PE-method can be illustrated by an example of the simplest flow system. Such system, with the point of view of its material structure, contains two half-space – of a high-and low-pressure – filled with a homogeneous fluid at equal temperature. The half-spaces are connected between itself with the flow element in the kind of a straight pipe with circular cross-section.

At traditional approach to a solving of the gas dynamics problems the specialists accept the above described sketch as a scheme for the computation. Taking the heat-transfer properties of the fluid as well as the geometrical and micro geometrical parameters of the flow element into account, they are carrying out the computation of the gas stream parameters in the pipe itself using the empirical coefficients for the energy loss along the pipe and its inlet and outlet. They suppose that the combination of a pressure drop between the fluid half-spaces with a flow element itself is sufficient condition for an originating of a fluid flow through the flow element.

In contrast to the traditional approach, PE-method envisages a composing of the General HydroDynamic structure of the flow system side by side with its material structure. The simplest General HydroDynamic structure contains in itself the following components:
- the high-pressure half-space – far from and near the pipe inlet; the fluid flow arises in the far zone and the fluid velocity increases sufficiently up to a quantity, equalled to the stream velocity at the pipe inlet in the near zone;
- the fluid stream in the pipe inlet part;
- the fluid stream in the pipe transitional part;
- the low-pressure half-space – far from and near the pipe outlet; in the near zone the fluid flow forms jet; in one's turn, the jet contains the jet kernel and around it – zone of a throwing out of the weight-flow.

In the General HydroDynamic structure it is necessary to ascertain strongly the boundary between the flow system and surrounding medium. In the considered case such boundary is in the pipe end section. The modern form of Saint-Venant – Wantzel's formula [3] determines the gas stream velocity just in the pipe end section. From the point of view of the interaction dynamics of bodies and mediums the contact interaction of the considered two fluid half-spaces takes place or it begins to take place also just in an outlet section of the pipe. On this boundary the fluid masses are leaving the flow system and carrying away the momentum together with itself in accordance with the second law of mechanics. At the same time, on this boundary the out-flowing fluid masses are inculcated in the low-pressure surrounding medium in accordance with the third law of mechanics. In the first of these cases the throwing out of momentum out of jet is accompanied by the oppositely directed impulse, and the phenomenon can be going on continuously. In the second case the inculcation phenomenon is the colliding and such phenomenon is single or cyclical. Therefore we have to elucidate the questions whether these two phenomena are compatible with each other and, if the compatibility is physically possible, what is its nature.

According to PE-method we consider only such mechanical system which possesses at least properties of inertness, elasticity and viscosity. These three properties form the base of a left part of the generalized equation of dynamics. The inertness (mass density) of a fluid is its capacity for an accumulating of a kinetic energy. The elasticity of a fluid is its capacity for an accumulating of a potential energy. The viscosity of a fluid is its capacity for transference of the mechanical energy impulse. The inertness and elasticity possess reversible nature and stipulate free transition of energy from its kinetic form to a potential, keeping an initial quantity of the energy. These two properties render the reaction for energy, applied from without. Therefore in the generalized equation of dynamics the terms, bound with inertness and elasticity, are called the inductive reactance and capacitive reactance correspondingly. The third property of fluid – a viscosity – stipulates permeability of fluid for the energy impulse, applied from without, and distribution of the energy potential correspondingly to a configuration of the walls, restricting a volume of the fluid. This property stipulates an irreversible nature of expense of the energy, applied from without, and one ensures a correspondence of the energy potential distribution to the every given flow system. The impulse of energy, applied from without, is spreaded at the sound speed in the given fluid and in direction, corresponding to the applied energy. In the generalized equation of dynamics the term, bound with a viscosity, is called the active resistance.

The above considered dynamic properties of real fluid are forming between itself combinations, pointing to the cyclical nature of perception of the applied energy by the fluid. Let $E$ is a dynamic module of elasticity of a fluid, $kgf/m^2$, $\mu$ is a dynamic viscosity of the fluid, $kgf \cdot s/m^2$, $\rho$ is a mass density of the fluid, $kgf \cdot s^2/m^4$, then

$E/\mu = \omega_{ve}$     is a frequency of a viscoelastic relaxation of the fluid, $1/s$;

$\mu/(A \cdot \rho) = \omega_{vi}$     is a frequency of a viscoinertial relaxation of the fluid, $1/s$; here $A$ is the stream cross-section area perpendicularly to the fluid motion direction, $m^2$;

$L\sqrt{E/\rho} = \omega_{ei}$     is a frequency of an elastoinertial relaxation of the fluid, $1/s$; here $L$ is the fluid stream length, $m$.

The adduced expressions show that a fluid, as a continuum, possesses three basic frequencies of relaxation depending on a relative quantity of the parameters, expressed by the above mentioned three basic properties and geometrical parameters of the flow system.

At the same time the adduced expressions show that the every mechanical system, including a flow system, perceives energy, applied to it, by the portions in accordance with basic frequencies of relaxation of the fluid. In other words, however energy was applied smooth, the mechanical system, including a flow system, perceives the applied energy cyclically. If energy, applied to a mechanical system, has also cyclical nature, then the forced oscillation of the system is determined by a ratio of the applied energy frequency and the system basic frequency and its harmonics. The adduced expressions show also that a viscosity determines one of the resonance frequencies of the forced oscillation of a fluid in contrast to the mechanical system, composed by the solid bodies and bound by Coulomb's friction. In the later the dry friction determines only the resonance magnitude but not the resonance frequency. This distinction allows explaining why two fluids, possessing the different viscosity, react by different ways under the equal cyclical action from without. For the low-viscous fluid the given cyclical action may be pre-resonant and for the high-viscous fluid the same action may turn out post-resonant. In one of these cases the fluid motion is controlled by viscoinertness, in the other case the fluid motion is controlled by viscoelasticity. In these two cases the equal action from without can lead to the exact contrary reaction of the fluid. Such example, known as Wesssenberg's effect, is adduced by R.B. Bird and C.F. Curtiss in their article [4].

Influence of the viscosity on frequency of a fluid relaxation consists in that the viscosity quantity determines a scale of reaction of the fluid to an action from without. In the low-viscous fluid the action is localized in the immediate proximity to the motion exciter, while in the high-viscous fluid the action from without is spreaded on the much more distance and carries along the much more mass of the fluid into motion. Thus the high-viscous fluid renders the heightened resistance to the action from without because not only of a large internal friction in itself, but also in consequence of a carrying along of the greater mass of the fluid into motion. Therefore the fluid viscosity, in contrast to dry friction, possesses more wide physical sense and renders an influence not only on the energy expense for the fluid motion, but also on the fluid motion trajectory.

A comparison of the above adduced expressions for fundamental frequencies of the fluid continuum shows that expressions for frequencies of the viscoelastic and viscoinertial relaxation is determinative and the expression for frequency of the elastoinertial relaxation is secondary. For example, a substitution $\lambda$ instead of $L$ reduces expression $L\sqrt{E/\rho} = \omega_{ei}$ to the form $f = c / \lambda$, that is to the simplest, with the point of view of physical sense, dependence of the oscillation frequency on a ratio of the sound velocity to a wave length, that is typical for a harmonic oscillation of system with dry friction or quite without it. In connection with the above stated it seems to be unattractive the near-physical mathematical manipulations around the non-Newton's fluid when the Newton's fluid motion theory is so far from being perfect.

Research of the motion on a base of PE-method envisages for the first stage the functional analysis of the every component of the General HydroDynamic structure of the flow system separately, and then the synthesis of the system on a base of unity of the cause and effect relationship, allowing to construct the integral physically adequate conceptual model of the fluid flow. In such approach the presence of a pressure drop and a flow system, is only necessary condition for origination of the fluid motion through the flow element (system). The sufficient condition for the motion consists in ascertaining of mechanism of interaction of the applied energy with fluid and the flow element as the self-oscillatory and self-organized process. Such properties as a reproducibility of the fluid stream parameters and the weight-flow characteristic of the flow element, system and its immutable correspondence with initial data allows to suppose the presence of the intrinsic voluntary negative servo feedback in the outflow system. An ascertaining of the feedback and its role in organization of the fluid motion on every section of the General HydroDynamic structure of the flow element is the essence of the given work. Author suppose the creating of the conceptual base for physically adequate description of the mainly gas medium motion as his principal problem, therefore the mathematical apparatus is reduced to minimum.

**To the jet kernel dynamics**
An ascertaining of the jet kernel dynamics is the key question of description of the fluid motion mechanism through the flow element, system. In the first place, it is stipulated that the jet kernel contacts both with the fluid stream on the end of the flow element and with medium of the low-pressure half-space. The jet kernel appears, as a rule, at a fluid outflow into the homogeneous with it a motionless medium. The out-flowing gas stream forms the greatest diversity of the submerged jet structures, among which it can be marked out the followings: a) low-speed, mainly laminar jet; b) subsonic, pressure jet; c) supersonic jet at the normal expansion, underexpansion, overexpansion; d) gas jet into vacuum. The liquid jet in atmosphere is an example of so called free jet.

In accordance with the results of the known experimental researches [5, 6] the initial section of an ax symmetrical submerged subsonic pressure gas jet contains in itself a kernel in the kind of a cone. The cone base coincides with the outlet cross-section of the flow element (pipe, nozzle). The kernel length (i.e. cone height) is about 4.5 diameter of the outlet cross-section of the flow element (i.e. cone base diameter). The flow velocity in the kernel volume is taken by constant and equal to the outflow velocity. Static pressure in the kernel volume is also taken by constant and equal to the surrounding atmosphere pressure at subsonic outflow.

More detailed information about the motion features of a gas and liquid submerged jets, obtained by means of the flow visualization methods, is presented by M.D. Van Dyke in his well-known Album [7]. Photographs in the Album show the jet structure under action both single impulse and at a steady-state flow. The other dignity of the Album is that the motion in many cases is presented by developing, as evolutional process, in the kind of series of the consecutive pictures. At the same time the modern methods of visualization allow showing only some features of the initial section kernel of a submerged jet because of an intensive turbulence, surrounding the kernel. The exception is only the jet part on the length about one its diameter from the outlet section of the flow element. As a whole, in spite of a large number of experimental researches and the developed theories, devoted to the jet dynamics, a problem of dynamics of the kernel of the submerged jet initial section and its role in the fluid motion process leaves unsolved.

Author of this work has obtained the key information on the given problem by means of observation with the naked eye the steam pressure jet, out-flowing out of a pipe with ~ 3 cm diameter into atmosphere at the temperature about zero degrees centigrade. At the beginning of observation the eye sees only chaotic motion of steam on the jet surface. After the expiry of the some minutes of the attentive observation the eye begins to fix the momentary separate steam puffs on the jet turbulized surface. These puffs replace frequently each other and they are evenly disposed on the surface, surrounding the kernel of the jet initial section. After appearance these puffs are passed away in the jet motion direction and they form turbulent jacket around the jet kernel. At further motion this turbulent structure forms so called basic section of the submerged jet.

The observation allows also ascertaining that a throwing out of the fluid masses out of the jet kernel is going on cyclically in the kind of a large number of the small separate jets.

A listening with the naked ear of the jet in combination with visual observation allows ascertaining that the sound generation is going on synchronously with the throwing out of a large number of the small jets from the jet kernel. A superposition of the sound of the small jets against each other forms a noise. Later on, the author was listening the gas submerged jets out of the flow elements of the different diameter and under action of the different pressure drop. These observations allow ascertaining the following: the more a flow element diameter the lower a noise tone, radiated by jet; the more a pressure drop the louder a noise, radiated by jet.

A comparison of the results of the known experimental researches and observations of the author shows that the kernel of the initial section of the submerged pressure jet is a body having quite definite form and dimensions. The body carries out the fluid weight-flow out of the flow element into surrounding atmosphere in the kind of a large number of the small jets, leaving the kernel cyclically. These small jets carry away with itself the momentum into surroundings and generate the corresponding elastic impulses of the recoil, directed inwards of the jet kernel. Thus the kernel as a body, possessing mass and elasticity, must accomplish the oscillations under action of the recoil impulses, directed inwards it. Fundamental frequency of the kernel, as a united whole, may be determined by known way.

Acoustic mass of the kernel, when its volume and density is known, is $m = \rho V$.

Acoustic hardness of the kernel, when, besides that, its sonic velocity and a base area is known, is $\kappa = \rho \cdot c^2 \cdot A^2 / V$.

Substitution of these expressions in the known formula of the cyclical frequency $\omega = \sqrt{k/m}$ gives $\omega = \sqrt{\rho \cdot c^2 \cdot A^2 / (V \cdot \rho \cdot V)}$.

If the kernel form is a cone, then its volume is $V = A \cdot h / 3$ (here $h = 4.5\,d$ and $d$ is the outlet diameter of the flow element) and then $\omega = c \cdot \sqrt{3 \cdot A/(Ah)} = c \cdot \sqrt{3/(4.5d)} = 2 \cdot c/(3d)$ and the oscillation frequency of the kernel as an united body is

$$f = 2 \cdot c / (2 \cdot \pi \cdot 3 \cdot d) \approx 0.106\,c/d = 0.106\sqrt{k \cdot g \cdot R \cdot T}/d.$$

The obtained expression confirms that a frequency of the noise, radiated by the kernel, is in reverse dependence on the flow element diameter and in direct dependence on the jet temperature.

At the same time, considering a degree of a packing of the above mentioned puffs on the jet surface, what was observed by the author, the kernel conical body must present by itself the system, consisting of masses, placed along the kernel axis by one after another and alternately colliding with each other. The every colliding is accompanied by a throwing out of the small jets, out-flowing into surroundings radially in the kind of a fan. In the result of such interaction a diameter of the masses is decreased. Longitudinal motion of the periodically colliding and subsequently decreasing masses forms the jet kernel body in the kind similar to a cone. One of the features of the above described structure of the submerged jet kernel is that its two last masses are very small and they can disappear periodically. If the masses, composing the kernel, possess an equal length, then its number, with taking into account of known results of experimental researches, must be from 8 to 10 in dependence on the oscillation moment. If length of one pair of the masses is equal to the flow element diameter, then the kernel length is from $4d$ to $5d$ with mean quantity $4.5d$. In connection with that the masses, composing the jet kernel, oscillate in opposite phases a frequency of the kernel oscillation as the multi-mass system must be twice as much a frequency of the kernel oscillation as the united whole, i.e.

$$f = 2 \cdot 0.106\,c/d = 0.212\sqrt{k \cdot g \cdot R \cdot T}/d. \qquad (1)$$

The conception of the submerged jet kernel structure as the multi-mass oscillatory system allows considering of a pressure in the kernel as a sum of static pressure, equaled to the surroundings pressure, and pressure, changing on harmonic in correspondence with a frequency of a throwing out of the fluid masses out of the jet kernel into surroundings,

$$p = p_h + \rho \cdot c \cdot v \cdot \sin^2 \omega \cdot t, \qquad (2)$$

here $v$ is the outflow velocity, determined for the gas stream by Saint-Venant – Wantzel's formula in its modern form [3].

Expression (2) may be wrote in the kind

$$p = \frac{\rho \cdot c^2}{k} + \frac{\rho \cdot c^2}{k} k \cdot M \cdot \sin^2 \omega \cdot t = \frac{\rho \cdot c^2}{k}\left(1 + k \cdot M \cdot \sin^2 \omega \cdot t\right) \text{ or } p = p_h\left(1 + k \cdot M \cdot \sin^2 \omega \cdot t\right) \qquad (3)$$

Thus, pressure in the kernel changes on harmonic within the limits

$$p_h \leq p \leq p_h\left(1 + k \cdot M\right). \qquad (4)$$

The maximum pressure correspond to the moment of the throwing out of the fluid masses out of the jet kernel into surroundings in the kind of the radial small jets and ratio

$$p_h / p_{\max} = 1/(1 + k \cdot M) \qquad (5)$$

allows determining a velocity of the small radial jets by means of Saint-Venant – Wantzel's formula in its initial form.

Expressions (2-4) show also that the noise level of the jet is proportional of its Mach number.
Equate the expression (5) with the well-known pressure critical ratio

$$\sigma_{cr} = p_h / p_0 = [2/(k+1)]^{k/(k-1)},$$

then after simple transformation we shall obtain

$$M_{cr} = (1 - \sigma_{cr})/(k \cdot \sigma_{cr}). \qquad (6)$$

For the air $k = 1.4$, $\sigma_{cr} = 0.528$ and according to expression (6)

$$M_{cr} = (1 - 0.528)/(1.4 \cdot 0.528) \approx 0.64.$$

The obtained result shows that when the jet outflow velocity $M \geq 0.64$ then a throwing out of the fluid masses out of the jet kernel corresponds to supercritical condition. An outflow of a jet under such condition will be accompanied by radiating of the shock waves and, accordingly, by the heightened noise. The radiation direction of the noise maximum intensity is determined by correlation of the outflow velocities of the main jet and the small jets, out-flowing out of it. Width of a wave front of the noise maximum intensity is commensurable with the jet kernel length.

Fig. 1 shows the diagram a) of the impulsive outflow of the radial small jets out of the main jet kernel into surroundings and the diagram b) of an action of the wave recoil impulses. Designation of zones and sections of the flow General HydroDynamic structure on fig. 1 and below are adopted from our article [1].

Diagrams on fig. 1 explain that the jet kernel has, in reality, the stepped form and its cross-section is decreased after every row of the impulsive out-flowing of the radial small jets out of the main jet. Last two masses – in a top of the jet kernel - exists during only half period of oscillation of the kernel.

Fig. 2 shows the simplified diagram of a disposition of the kernel of a supersonic submerged jet, having the extension section in surrounding atmosphere.

In the whole, diagrams on fig. 1, 2 explain a forming of the recoil summary impulse, directed against motion of the gas stream. Consecutive throwing out of the jet masses along the kernel length transforms it in a megaphone, concentrating the radial impulses from its lateral surface in the united longitudinal impulse. If the jet outflow velocity is $M > 0.64$, then this megaphone radiates against motion of the gas stream the shock waves. Such running waves move against flow both on supersonic and subsonic sections of the gas stream.

In certain sense the described phenomenon is similar to action of the explosive charge with conical hollow, although these two processes are quite opposite in a direction of its proceeding. In the first of these compared cases the motion of the fluid stream into the megaphone hollow is accompanied by a throwing out of the stream weight-flow through the lateral surface of the megaphone; the latter concentrates the recoil impulses and forms one strong stream of the wave impulses, directed along its axis. In the second case a motion of the detonation wave along an axis of the conical hollow is accompanied by a throwing in the matter to the hollow through its lateral surface; the latter concentrates the moving matter and forms one strong jet, directed along its axis. A distinctive feature of the first case is that the described process is self-organized by the submerged jet motion. The other feature of the first case is that the small radial jets, out-flowing out of the jet kernel, are inculcated into the surrounding atmosphere and carry the latter along direction of the main jet motion; this phenomenon is named the ejection.

The examples of the flow visualization of the submerged jet motion, presented in above mentioned Album [7], show that such jet motion can be also accompanied by forming of the spiral ring-shaped vortex (SRV) structures on a boundary of interaction of the submerged jet with homogeneous with it surrounding medium.

Fig. 3 presents schematic above mentioned jet. Such SRV structures appear not only at a steady flow of the submerged jet, but and at the impulsive throwing out of a fluid into homogeneous with it surrounding medium. At a relative motion of two fluids, having a plate boundary, is formed the spiral-cylindrical vortex (SCV) structures, oriented across the relative motion direction of the contacting fluids. Such structures appear, in particular, in the track behind a bluff body, moving in the motionless fluid; they accompany also the free and forced convection. Wide distribution of the spiral vortex (SV) structures and likeness of its conduct allows choosing these structures in a special class, for which it is necessary to develop the dynamics generalized conception.

Features of SV structures are the followings:
- these structures appear mainly in zones of an inculcation of one fluid into the other and of the tangential contact interaction of the fluid layers;
- these structures ensure the continuity condition on the tangential interaction boundary of the fluid layers by way of transforming of the relative translational motion of the every fluid layers into a mutual angular motion;
- these structures limit a zone of the tangential interaction of the fluid layers according to its continual and boundary viscosity between its as well as its relative velocity and Archimedes force action;
- these structures appear, develop and degrade by way of evolution and are quite self-dependent, displaying the self-organization property.

The listed features of SV structures allow identifying the form of its motion on the stage of its developed existence with such transcendental curve as Euler (1744) – Cornu (1874) spiral – clotoid. The clotoid natural equation is $R = k / L$ and it means that the curvature radius $R$ is inversely proportional to the arc length $L$ of the curve. Clotoid has two special points: a beginning point at $R \to \infty$ and asymptotical point at $R \to 0$. Other feature of a clotoid is that its curvature radius is decreased very intensively on its two beginning coiles $L \approx 2.718$ relative length and further decreasing of its radius is going on very smoothly around the asymptotical point, forming so called a rotary kernel. The clotoid natural equation can be compared with the simple kinematical expression $R = v/\omega$, where $R$ is a rotation radius, $v$ is a translational motion velocity and $\omega$ is an angular frequency the rotary motion. When $\omega \to 0$, then $R \to \infty$ the motion will be translational, and when $\omega \to \infty$, then $R \to 0$ and the motion will be rotary. Thus a clotoid is a function, expressing a smooth transition of the translational motion into the rotary motion according to a simple kinematical correlation.

Fig. 4 shows a clotoid positive branch.

Fig. 5 shows two possible combinations of two branches of the clotoid.

Fig. 6 shows the combination of three clotoides, forming the successive stages of development of the mushroom-shaped hydrodynamic structure and subsequent formation of SRV structure in homogeneous motionless fluid. Feature of SV structures is that these structures are dynamically unbalanced. It is connected with that a rolling of the structure external layers is accompanied by a decreasing of its thickness and, correspondingly, by an increasing of its rotation frequency and, as a consequence, by an increasing of centrifugal forces acting in the kernel, revolved around asymptotical point. Under action of these centrifugal forces a fluid in the kernel cavity is thrown off to its quick-revolved surface and static pressure in the cavity is decreased. Thus the kernel vacuums itself. On the kernel surface its centrifugal forces is counterpoised by the compression forces of the SV structure external layers, therefore the kernel cross-section diameter is always remained invariable. In contrast to it, the rotation velocity of the kernel is continually increased and therefore pressure within the kernel is continually decreased at constant pressure in atmosphere, surrounding SV structure. Thus pressure drop on the kernel surface is continually increased and, in the end, reaches the quantity, corresponding to the buckling. In this situation the kernel is bent and then it is disintegrated and destroys SV structure.

Fig. 7 presents the development stages of SCV structures, when two homogeneous fluids *a* and *b* move in the exact opposite directions. Initial boundary of these fluids is a plane, designated by a straight line 1. Then the line 1 transforms into sinusoid 2. Such form of the contact surface corresponds to the mutual buckling of the boundary layers under action of the longitudinal forces its contact interaction by friction. An increasing of the sinusoid magnitude leads, in beginning, to a structure 3 – trohoid, then to a structure 4 – wave with the white-cap and further to a developing of SCV structure 5, formed by the conjugate clotoides. A step of SCV structure 5 is equal to the wave-length of sinusoid 2 and the asymptotical points of the clotoides are disposed on one line. The boundary layer thickness of each fluid is ~ 0.577 wave-lengths. After that SCV structure loses the longitudinal stability, its elements are shifted in transversal direction as on a chess-board. Now the kernels of the vortexes disintegrate, absorb the neighbors and form with the neighbors the pair. In particular, the running waves on the water surface under the wind action are the clotoid embryos.

Fig. 8 shows the conjugate of clotoides, reproducing the well-known vortex street in a track behind a bluff body, moved in a motionless fluid.

Fig. 9 explains a difference in the scale and nature motion of the high-viscous (*a*) and low-viscous (*b*) fluids under a rotating its with shaft, when the shaft axis coincides with the cylindrical vessel axis. A view from above is showed here. The vessel diameter is equal to a diagonal of the square in which is inscribed one branch of the clotoid. The shaft diameter is equal to the clotoid kernel diameter. Therefore a ratio of the vessel diameter to the shaft diameter is equal to 7. The fluids under action of the shaft rotation are aspired to move along the clotoid, as it is showed on fig. 6 – 8. The high-viscous fluid motion is determined by two beginning branches of the clotoid. The angle between the beginning of the clotoid and the vessel cylindrical surface is 45° and one is decreased practically to a zero on the end of the clotoid second branch (close by the shaft surface). The components of the motive force in immediate proximity to the vessel wall are directed along the tangent to the vessel wall and along the vessel radius and ones are equal by quantity. The tangential force excites a rotary movement of the high-viscous fluid (along the vessel wall). At the same time, the radial force excites a straight-line movement in direction to the vessel axis (to the rotation centre). When the shaft rotation frequency is low and the centrifugal forces are insignificant, then the radial force provokes the circular hollow on the high-viscous fluid free surface along the vessel wall and a hillock around shaft.

The low-viscous fluid motion is determined by the clotoid branches in immediate proximity to its asymptotical point. Now the vessel diameter is equal to the clotoid kernel diameter and the shaft diameter, as before, is less of it in 7 times. In this case the angle between the clotoid branches and vessel wall is quite insignificant. Therefore the radial component of motive force is also quite insignificant. Under the shaft rotation the centrifugal forces surpass the radial component of the motive forces and throw off the low-viscous fluid to the vessel wall, forming an ax symmetrical hollow on the fluid free surface around the rotating shaft. In that is the physical sense of Weissenberg's experiment in contrast to the pretentious novelties of respectable mathematization experts [4] by means of polymeric noodles.

The above stated description of a forming of SV structures quite corresponds to the results of a welding of the plastic metals by means of the explosion, adduced in [8]. In this case the sinusoidal and SV structures are formed along the joint surface of the welded metallic plates, when the explosive detonation velocity exceeds the sound velocity in the metal and the shock wave intensity exceeds the limit of plasticity of the metal.

However, the nature gives itself the very available example of a forming of SV structures for every one who wishes in the kind of the banks of the stratus clouds.

Now, after the qualitative analysis of some SV structures, the consideration of the submerged jet dynamics can be continued in accordance with scheme on fig. 3. The feature of such jet is in the following. When the fluid stream moves through the flow element, then its near-wall velocity is less than the near-axis velocity in consequence of a friction. When the stream flows out of the flow element into the surroundings, then the friction quantity and its character is considerably changed. Therefore the velocity profile on the initial section of jet becomes level. In the result the velocity of the external layers of jet is some increased and the near-axis velocity is some decreased on the length ~ 1.5 diameters from the end of a flow element. The longitudinal shear of the external layers of jet relatively its near-axis part is a cause of a possible forming of the embryos of SRV structures in the kind of the sinusoidal profile. At the same time, if the jet velocity is very small in comparison with the sonic velocity (and $M \ll 0.64$) and the motionless surrounding atmosphere is homogeneous with jet, then a forming of the above described SRV structures on its boundary is quite possible. A subsequent formation of SRV structures along the initial section of jet leads to a subsequent dispersion of the jet weight-flow in surroundings. Every SRV structure radiates into the jet kernel the elastic impulse. The jet kernel pulsates, it possesses the megaphone form and it radiates the elastic waves against the fluid flow. An out-flowing of the free jets is also accompanied by appearance of a stream of the elastic waves against the stream motion. An example of the gas free jet is an out-flowing of the gas stream into vacuum. In a vacuum the gas jet is already supersonic. An influence of the pressure drop is that the dimensions of the jet expanding section is the more the higher the pressure drop. Feature of the jet is that the jet kernel in it is absent an the jet expansion is going on up to its compactness loss, when the gas particles begin to move by inertia, independently each other. A generating of the elastic waves against the stream motion is going on in the end cross-section of the flow element.

A liquid free jet is differed from the gas free jet by an absence of the expansion section. At the same time, the liquid free jet like the gas free jet not possesses the jet kernel and a generating of the elastic waves against the stream motion is going on like the gas free jet in the end cross-section of the flow element.

In conclusion of the section of the given article it is necessary to elucidate the questions on the nature of the wave impulses, radiated by the kernel of the submerged jet initial section, and its difference of the nature of the wave impulses, accompanying the hydraulic shock phenomenon. In the case of the hydraulic shock the sudden braking of the liquid stream in a flow element (pipe) is accompanied by a transition of the stream motion kinetic energy to the potential energy in the kind of the extra static pressure. This transition moves from the stream braking place – with the velocity determined by elasticity of the fluid and the pipe wall – against the stream flow and increases the static pressure in the pipe before the braking on the quantity of the extra static pressure. The summary pressure is also the static and one acts on the pipe wall. The summary pressure can be measured by the ordinary manometer. The hydraulic shock can also arise at the sudden acceleration of a fluid stream. At the uniform steady motion of the liquid stream the hydraulic shock is not arises. In contrast to it, the stream (both gas and liquid) outflow out of the flow element is always accompanied by a radiating of the wave impulses, directed against the stream flow. Pressure in the radiation front has the vector nature, one is not static, one not acts on the pipe wall and one cannot be measured by manometer. The elastic waves, running from the flow element end, are the direct consequence of the second and the third laws of mechanics in the kind of a reply to a throwing off of the fluid masses out of the flow system. The motion velocity of these elastic waves is determined only by the fluid sound velocity, in contrast to the hydraulic shock. Therefore the expression (1) is equally true for both liquid and gas. These waves put over the mean velocity of the fluid motion only the oscillatory component in a longitudinal direction.

**To the dynamics of the transitional and inlet parts of the flow element**
In the all diversity of the constructive features a structure of the simplest flow element can be presented in the kind of its transitional and inlet parts. The transitional part in the simplest case can be presented in the kind of the straight pipe with the circular cross-section and with length from zero to very large quantities. The most general geometrical characteristics of a pipe are it's the dimensionless caliber length, ($L / D$), and its wall dimensionless relative roughness. Geometrical diversity of the inlet part can be presented by the profiles from the point-shaped (an orifice in the thin wall or a straight cut of the pipe inlet part) and the curvilinear. The most general geometrical parameters of the inlet part are its relative area of the flow cross-section and its profile relative roughness.

The basic practical problem during development of a flow element, system is that to ensure of the most weight-flow of the flow element, system in combination with the acceptable simplicity of the flow element construction. In such statement of the problem the flow element is considered as a device for an ensuring of a quite definite weight-flow.

The essence of a physical aspect of this problem is that to elucidate the role of its structural elements in the mechanism of a forming of the fluid motion and the influence of the geometrical features of the flow element on the trajectory and the intensity of the fluid motion. From these positions the role of the flow element transitional part is the following:
- the transitional part diameter determines the oscillatory frequency of the kernel of the jet outflowing out of the flow element;
- the transitional part is a waveguide for the elastic waves, running from the jet kernel, as a megaphone generator-radiator, to the high-pressure half-space;
- the transitional part length with the taking into account of the conditions on its ends determines its oscillation fundamental frequency and its harmonics;
- the transitional part ~ 3 calibers length transforms the concave front of the wave, running from the jet kernel, in the plate front.

Between the length of the flow element transitional part in the kind of a pipe and the jet initial section kernel is the following correspondence. The fundamental frequency of the fluid column in the pipe, acoustically opened on both ends, is $f_0 = c / (2L)$ and its harmonics is $f_1 = c / L$, $f_2 = 3c / (2L)$,... When the frequency of the oscillations, generating out of the jet kernel, in accordance with the expression (1), is $f = 0.212 c/D$, then the pipe resonant lengths is $L_0 = L / D = 2.359$, $L_1 = 4.717$, $L_2 = 7.075$, ... The resonance on the pipe fundamental frequency in a combination with the equality of the fluid impedance on the pipe inlet and the high-pressure half-space ensures the free motion of the running waves from the pipe end to its inlet and in that way one ensures the minimum hydraulic resistance for the fluid stream in pipe in accordance with the experimental results, adduced, for example, in [9]. When the pipe has the sharp changes of its diameter, then the running wave is partially reflected from such places. The reflected (reverse) wave returns to the generator of the oscillations and forms the coincident wave with the forward wave. In this case the wave losses of the energy of the fluid stream are determined by means of the standing-wave ratio and the traveling-wave ratio as it is accepted in the theory of the oscillatory system with the distributed parameters.

Perfection of every pipe as a waveguide can be valued by means of the shock transfer coefficient

$$\eta = 4\rho_1 \cdot c_1 \cdot A_1 \cdot \rho_2 \cdot c_2 \cdot A_2 / (\rho_1 \cdot c_1 \cdot A_1 + \rho_2 \cdot c_2 \cdot A_2)^2,$$

where $\rho_1 c_1 A_1$ and $\rho_2 c_2 A_2$ is the wave resistance of the every of two contacting mediums and $\rho$, $c$, $A$ is the mass density, sonic velocity and the contact area, correspondingly. When the specific wave resistance, $\rho c$ of two contacting mediums is equaled, then the previous expression accepts a form

$$\eta = 4 A_1 \cdot A_2 / (A_1 + A_2)^2.$$

As it is applied to the flow system dynamics, $A_1$ is the pipe cross-section area and $A_2$ is the cross-section area of the high-pressure half-space, where flow is began.

When $A_1 = A_2$, then $\eta = 1$. It means that the only forward running wave moves without a reflection in the outflow system. The wave losses are absent in this case.

When $A_1 < A_2$, then $0 < \eta < 1$. It means that the part of the running wave energy is reflected from the place of the change of the stream cross-section area with a forming of the coincident wave. Now the wave losses are determined by the standing-wave ratio or the traveling-wave ratio.

When $A_1 \gg A_2$, then $\eta > 0$. It means that the part of the running wave energy is carried out of the flow element, system into the high-pressure half-space and one ensures a forming of the inflow zone in front the pipe.

At the smooth change of a cross-section of the flow element transitional part the plate front of the running wave transforms to a sphere-shaped – concave or convex – in dependence on a convergence or divergence of the transitional part. The combination of the plate and sphere-shaped fronts in the some cross-section or zone of the transitional and inlet parts forms so called hydrodynamic (HD) lens. Such HD-lens is dispersal for the running waves and in the same time one is gathered for the fluid flow and vice versa. The surfaces of the HD-lens are contacting with the streamlined surfaces of the flow element by the normal. The HD-lens surface area determines the stream cross-section area in contrast to the traditional approach, when the every cross-section is accepted by a plate. A ratio of the area of the lens surfaces is its so called the multiplication coefficient and one determines a degree of the wave front form: concave, convex or plate and in this way a degree of the fluid stream expansion or contraction.

The conditions of a going out of the running wave out of the pipe into the high-pressure half-space determine the energy losses in this zone. The quantity of the losses depends on the form and the intensity of the front of the running wave.

Fig. 10 presents the profile schemes of the well-known inlets of the simple flow elements: a) an orifice in the thin wall; b) Borda's nozzle (the sharpest inlet); c) sharp inlet. The white arrows indicate the motion direction of the fronts of the running waves. The filled arrows indicate the fluid flow direction. The designation of the zones of the flow hydrodynamic structure is adopted from our article [1]. Here letter *b* designates the HD-lens zone. The all fronts of the running waves are constructed by means of the united method – the circumference evolvent. In the all cases the fluid flow trajectories form the fields, orthogonally coupled with the fronts of the running waves. Thus the pressure drop is the motive force, and the fronts of the running wave determine the direction and intensity of the fluid flow. The schemes a), b), c) on fig. 10 show that, in spite of the constructive differences of the flow elements, its HD-lenses are identical and their multiplication coefficient is 4. In this case the fluid, inflowing into the flow element (orifice, nozzle, pipe), has the excessive high speed. Therefore the fluid flow is accompanied by its separating from the flow element wall.

Fig. 11 shows the profile of: a) conical inlet; b) circular inlet. Two HD-lenses ensure the summary multiplication coefficient of the conical inlet, equaled to ~ 5.62. Such inlet ensures the almost unseparated flow. The excessive large multiplication coefficient of the circular inlet, equaled to ~ 9.68, ensures the unseparating flow in the flow element inlet part, but one increases significally the friction losses because of the excessive length of its profile and thus one increases the inlet summary losses.

Fig 12 presents a profile scheme of the flow element inlet part, constructed by means of the circumference evolvent, when the circumference radius is 0.47 of the pipe radius. In this case the multiplication coefficient is equal to 8 and this profile ensures the least losses. The geometrical parameters of this profile, constructed on the base of the united curve, quite corresponds to the particular recommendation for the most perfect inlet, what was found at the close of the 19$^{th}$ century [10]. The analogical solution is used for a profiling of the outlet part of the musical wind-instruments. Their outlet part contains two parts: the conical slightly diverging part and the curvilinear intensively diverging part. This second part is profiled in the kind of a circumference evolvent. The combination these two parts ensures the minimum wave losses and therefore one ensures the greatest intensity and uniformity of the radiated sound. The friction losses, in contrast to the industrial flow system, in the wind-instrument are practically absent owing to the air insignificant weight-flow. In the industrial flow system the greatest weight-flow can be reached on base of a minimization of the combination of the friction losses with the wave losses. A profiling of the flow element inlet part, as it is showed on fig. 12, ensures the necessary compromise: a significant diminution of the evolute circumference radius of the flow element, in comparison with the wind-instruments, increases the wave losses insignificantly and the significant shortening of the inlet profile significantly decreases the friction losses.

Thus the well-known results of the experimental research of the industrial flow elements and the features of the musical wind-instruments obtain the quite sufficient physical explain on a base of the conception about the fluid flow as a process of the contact interaction of a fluid with the flow element wall under the wave mechanism of the flow regulation and under action of the pressure drop.

**To a dynamics of the fluid motion through the flow element**

The above considered features of the fluid motion dynamics in the every section, part and zone from the high-pressure half-space to the low-pressure half-space through the flow element allow presenting the fluid flow in the kind of the integral dynamical structure and formulating the general principle of the motion.

The principle is that the pressure drop, applied to a fluid in the flow element, system, is a necessary condition for the motion appearance, and the reverse stream of the elastic waves, running from an end of the flow element, system, is the sufficient condition for a forming of the motion structure; the interaction of the pressure drop with the reverse stream of the running waves completely determines the flow dynamics correspondingly to the flow element constructive features.

Fig. 14 presents: a) scheme of the fluid stream motion through the flow element; b) scheme of a generation and motion of the fronts of the elastic running waves against the fluid motion. These schemes present the subsonic outflow.

Fig. 15 presents the simplified scheme of a generating of the elastic waves, of its motion in the transitional part of the flow element and of a radiating of the waves into the high-pressure half-space. This scheme corresponds to the supersonic outflow of the gas stream under action of the hypercritical pressure drop [1: Fig. 1, scheme 14].

Schemes on fig. 14, 15 contain the General HydroDynamic structure of the flow [1] and present the fluid motion as the integral, self-organized, self-oscillating system, including the all components of the self-regulation: the generation and focusing of the elastic running waves (on the flow element end), the transference of the wave energy by the waveguide (in the flow element transitional part), the correspondence of the transitional stream impedance with the high-pressure half-space impedance (in the inlet part of the flow element) and a radiation of the running waves in the kind of the near and far fields (in the high-pressure half-space). The cyclical entrancing and spreading of the wave fronts into the high-pressure half-space accomplishes the functions of a vectorization of the initial chaotic motion of the fluid particles and the intensification of its motion in the direction

4to the flow element inlet. The intensity of the wave fronts always corresponds to the pressure drop quantity and the configuration of the wave fronts corresponds to the flow element profile.

Thus the regulating of the flow direction and its intensity is ensured by the internal voluntary servo feedback. Just the servo feedback ensures the reproducibility of the flow structures, the state and motion parameters of the fluid stream and the weight-flow characteristic of the flow element.

**Discussion of results**

From the stated in given article position the presence of the fluid as the incompressible and non-viscous substance leads to the physically empty mathematical manipulations and the posterior declarations that the nature is paradoxal, surprising, exotic, elusive. Such speculations are natural towards the end of the 20[th] century, when the computers and supercomputers ensure the practically unlimited possibilities for the physics mathematization [3, 8, 11, and 12].

On the contrary, the approach, stated in the given paper, not only increases the possibility of the hydrodynamic calculation by that way of the addition with the new laws, but one opens the new possibilities by means of a using of the mathematical apparatus of the acoustics and the alternate current for a modeling of the fluid flow.

The indissoluble connection of the mass density and elasticity of the fluid with the wave nature of the motion self-organization under action of the motive force and a unity of a motion process possess the direct connection with L. de Broglie's (1923) and then N. Bohr's, H. Kramers` and J. Slater's (1924) statement that the all bodies and mediums in nature possess both the corpuscle and wave properties simultaneously. Conformably to the solid, liquid, gas and plasma dynamics this principle envisages the presence in the bodies and mediums, as a minimum, the mass (inert) density and the elasticity (continual and contact). The presence of these properties determines mass-wave dualism of the motion process. The first signal on the motion dual nature was arose in the 18[th] century, when Torricelli's formula was wrote rightly $v^2 = 2gh$. The spelling of this formula in the kind

$$h = \frac{1}{g} \cdot \frac{v^2}{2}$$

testifies that the free fall velocity is changed on the harmonic law and its mean square quantity corresponds to the fall height. In this case $h$ is simultaneously both the fall height and a difference of the potentials of the gravitation field. G. Coriolis (1829) generalized C. Huygens` (1669) and G. Leibniz` (1686) research results and gave the right expression of "live force" (viribus motus) in the kind $m\,v^2/2$.

In parallel of these researches Galilei and Huygens studied the pendulum and string oscillation. Sir Rayleigh [13] has developed the theory of the oscillations. In the frame of this theory Coriolis` spelling for the uniform forward motion can be presented in the kind

$$(1/2)mv^2 \to mv_0^2 \sin^2 \omega \cdot t \to m\frac{v^2}{2}.$$

In the words: the kinetic energy quantity of a body, moving uniformly and rectilinear, is equaled to the product of the body mass and the mean square velocity, changing by the harmonic law. On the face of it, such formulation seems to be discrepant. But the wave theory of the shock of the solid bodies, developed in [14], discloses the wave nature of the interaction of the solid bodies at the direct colliding. After a colliding the bodies can move uniformly and rectilinear, but the kinetic energy quantity of these bodies has been determined by the wave nature of its contact interaction, therefore one is expressed by means of the mean square velocity, acquired in a process of the colliding. The energy of the elastic deformation of the solid bodies is determined, in accordance to [15], by the product of the rigidity and the mean square elastic deformation.

Thus, the experience testifies on the wave nature of the contact interaction of bodies and mediums and their further free motion can be uniform and rectilinear. The considered examples possess the nature of principle, testifying that the wave mechanism of the energy motion in bodies and mediums is to be universal and one determines direction and intensity of a motion not only at its contact interaction, but also under action of such force field as the gravitational field.

**Final remarks**

The above stated approach to a solution of problems of the fluid dynamics by means of Physical Ensemble method is essentially distinguished from the multidisciplinary approach, arose in the second half of the 20[th] century, but one quite correspond to Aristotle's principle: "there is no motion without physical body."

**Acknowledgements**

_______________________________

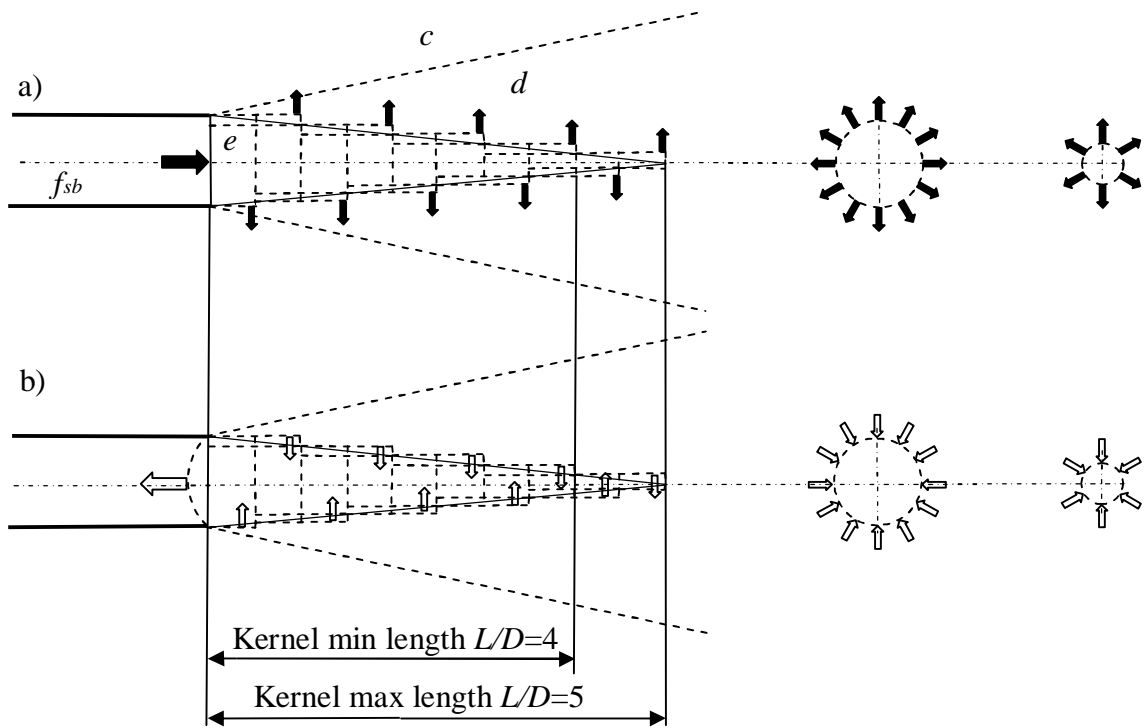

Fig. 1. The submerged subsonic pressure gas jet out of the pipe end part: a) scheme of impulsive throwing out of a gas in the kind of radial elementary jets out of jet kernel into surrounding atmosphere; b) scheme of a forming of the recoil wave impulses

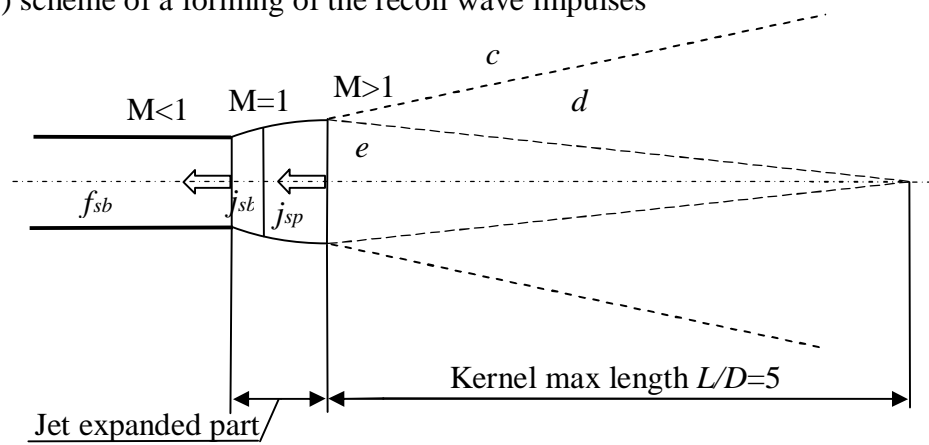

Fig. 2. The submerged supersonic gas jet out of the pipe end part under hypercritical pressure drop; simplified scheme [1: Fig.1, scheme 14]

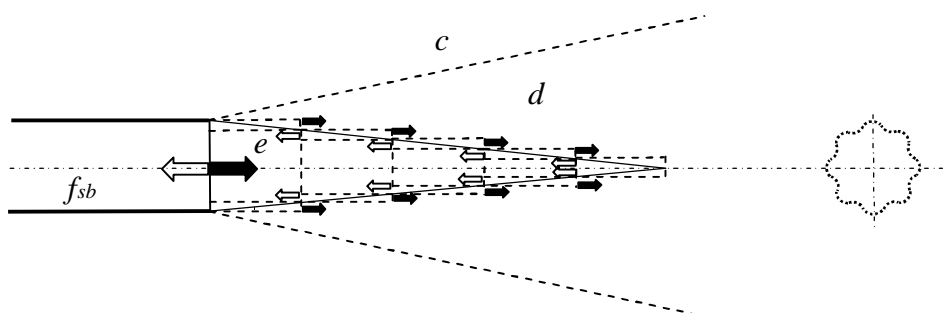

Fig. 3. The submerged subsonic low-pressure gas jet out of the pipe end part: a combined scheme of impulsive throwing out of a gas in the kind of ring-shaped vortexes out of jet kernel into surrounding atmosphere and forming of the recoil wave impulses

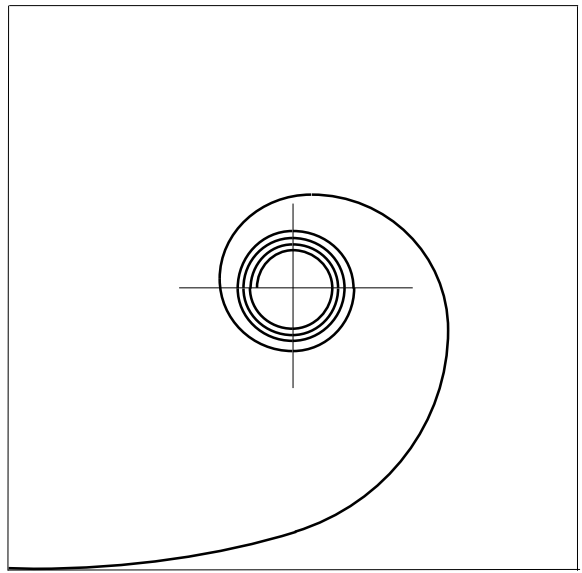

Fig. 4. The clotoid – positive branch

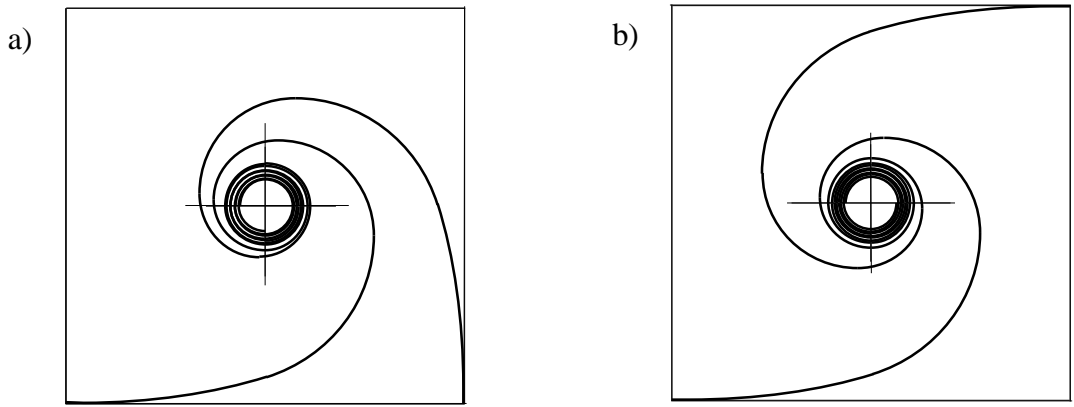

Fig. 5. Combinations of two clotoid

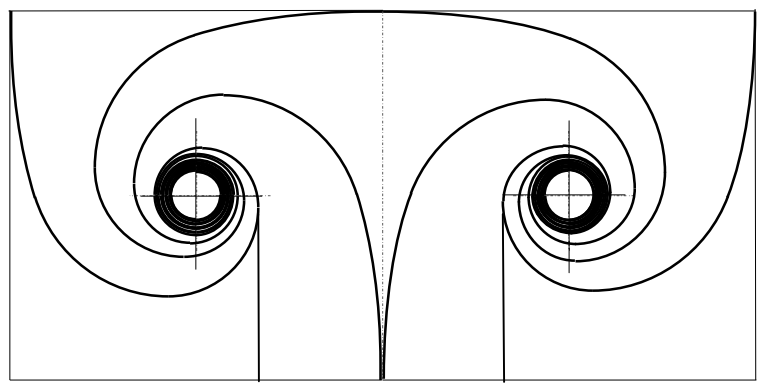

Fig. 6. Combination of three clotoids: transformation of a mushroom-shaped structure into a spiral ring-shaped vortex

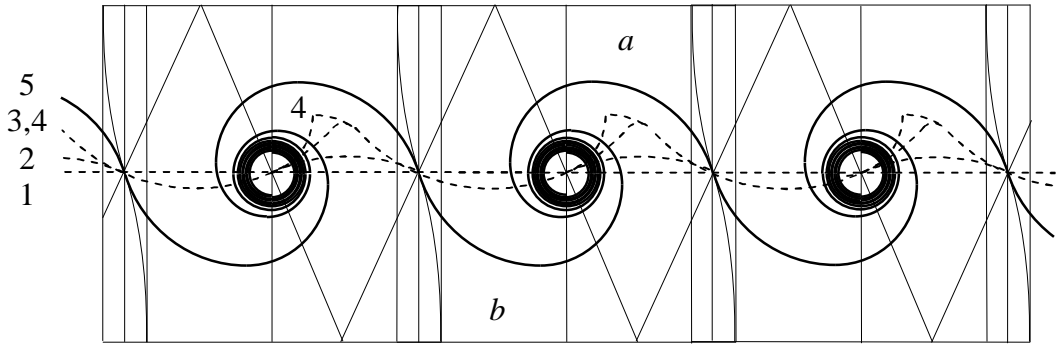

Fig. 7. Evolution of the profile form of a contact boundary of two homogeneous fluids *a* and *b* at its relative horizontal motion: 1 – initial straight line; 2 – sinusoid; 3 – trohoid; 4 – wave with crest and white-cap; 5 – combination of clotoids

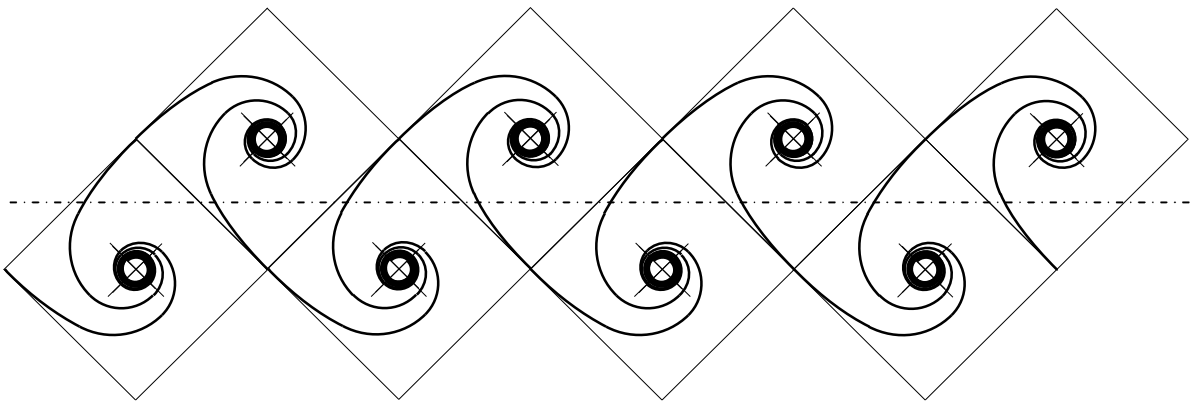

Fig. 8. The track behind a bluff body is constructed by means of a clotoid

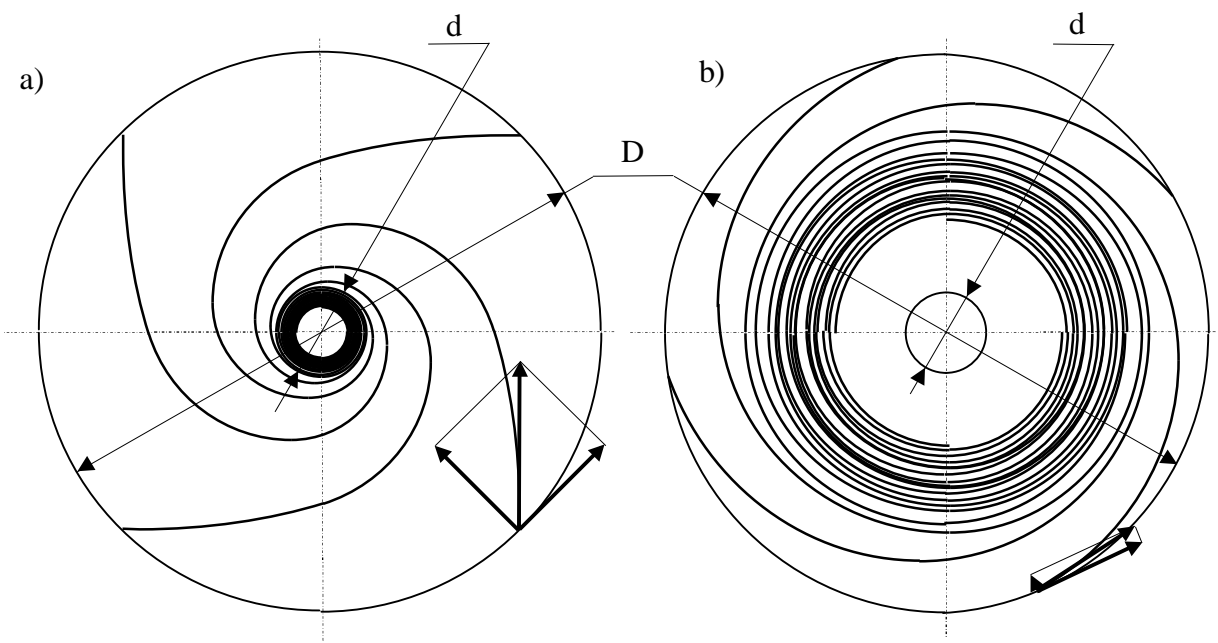

Fig. 9. The fluid motion in the cylindrical vessel under action of a rotating shaft that is coaxial to the vessel and that is rotated anticlockwise: a) high-viscous fluid; b) low-viscous fluid
(Notice the optical illusion: the right picture is seems smaller than the left one)

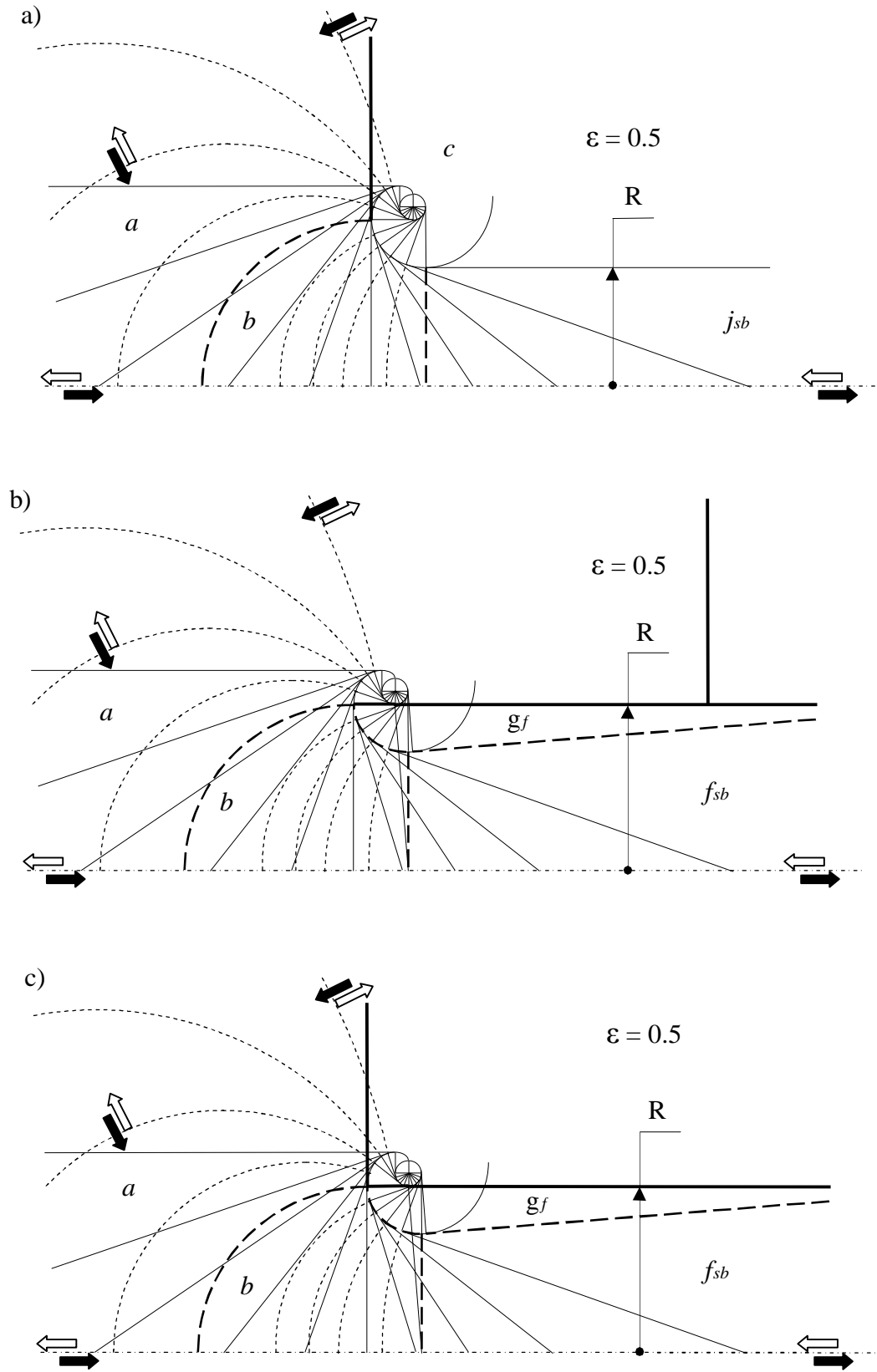

Fig.10. The fronts of running elastic waves and forming of a fluid stream in the inlet zone:
a) outflow through the orifice in a thin wall; b) the sharpest inlet – Borda`s nozzle; c) sharp inlet in a pipe; ➡ - fluid flow; ⇐ - direction of running elastic waves; *b* – hydrodynamic lens

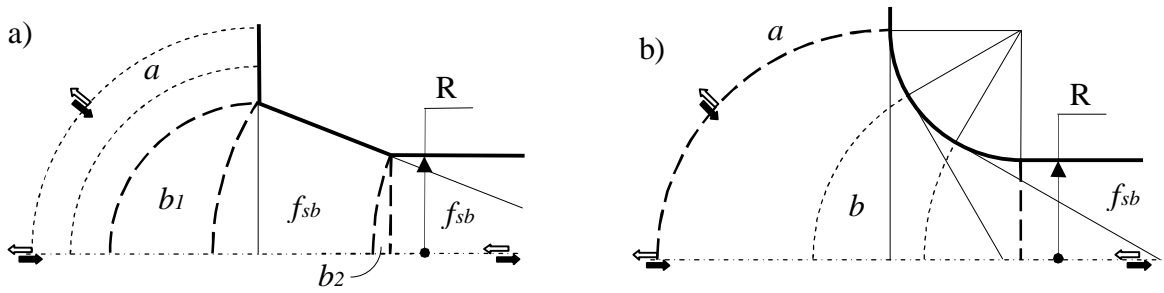

Fig.11. The fronts of running elastic waves and forming of a fluid stream in the smoothed inlet zone: a) conical inlet; b) circular inlet; ➡ - fluid flow; ⇐ - motion of running elastic waves; *b, b₁, b₂* – hydrodynamic lenses

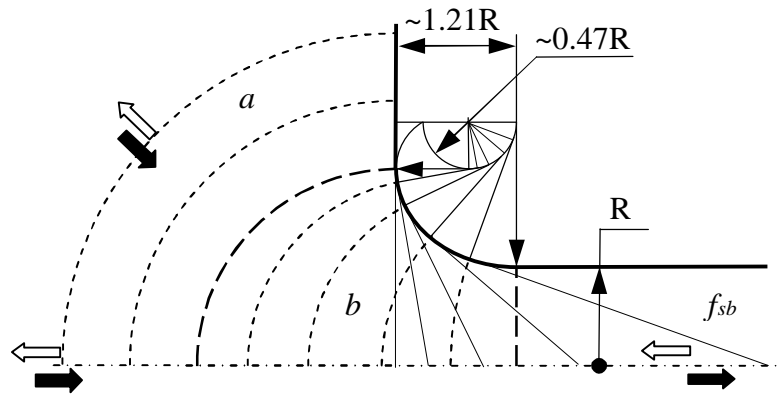

Fig.12. This evolvent is the most perfect inlet profile for the tank – pipe and tank – nozzle systems

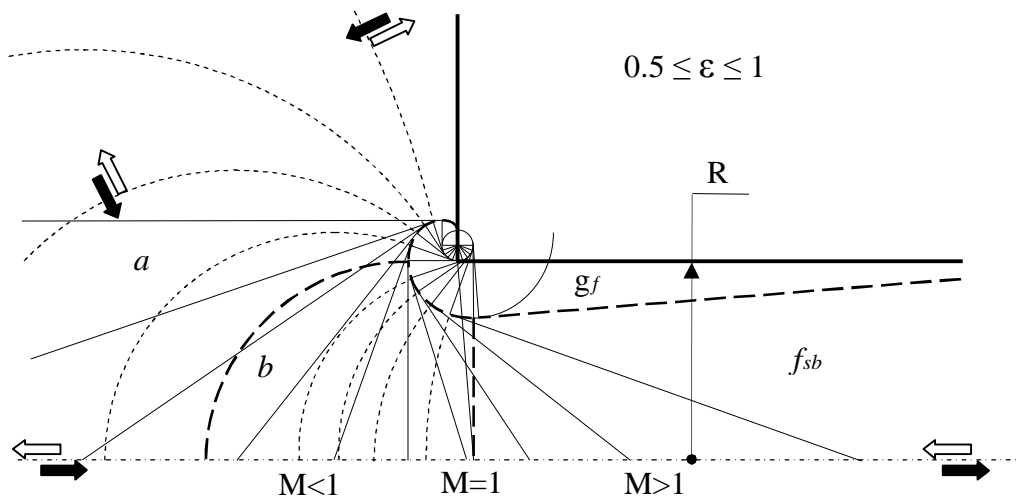

Fig.13. Forming of a fluid stream in the inlet zone of 9-caliber pipe at supercritical pressure drop [1: Fig.1, schemes 7 - 9]

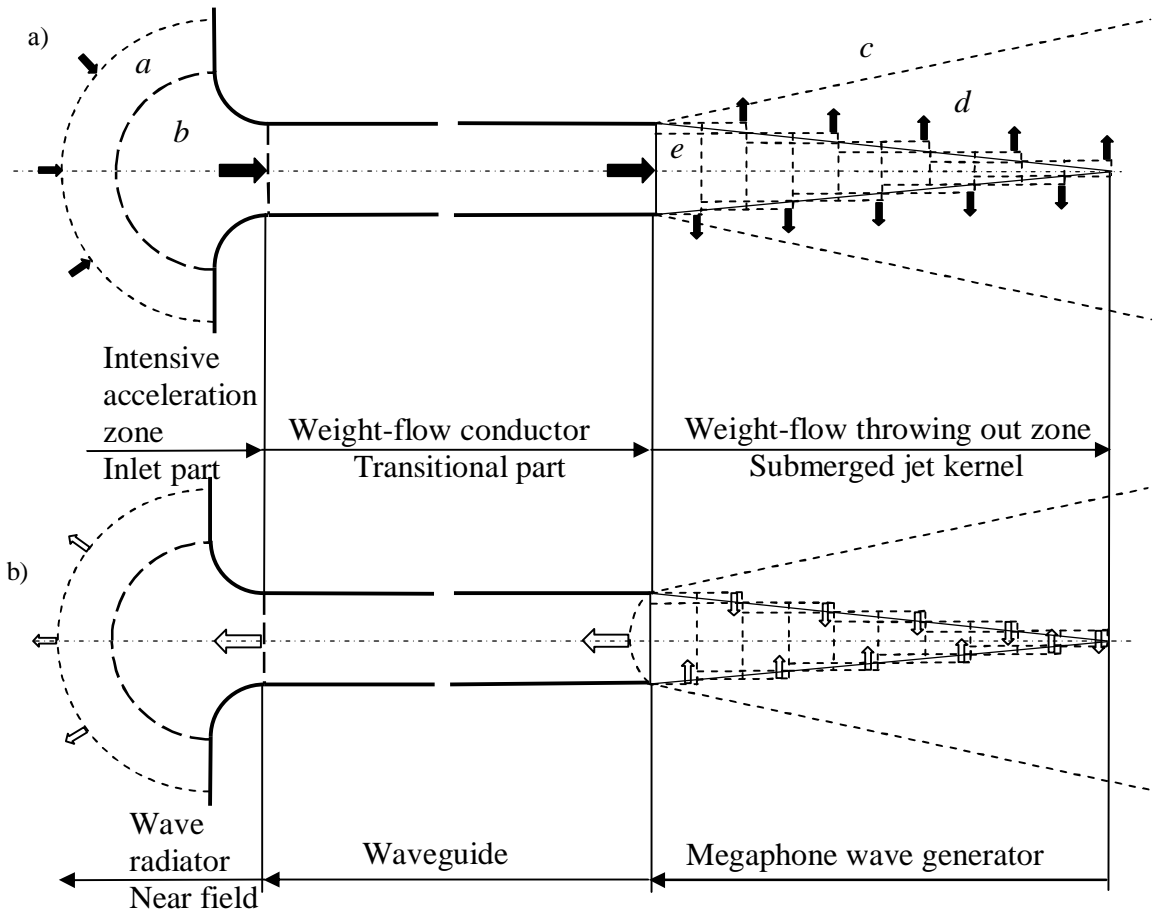

Fig.14. The fluid subsonic motion components: a) the fluid mass transference from the left to the right possesses a kinetic energy (outflow velocity is the function of pressure drop); b) the massless controlling energy transference from the right to the left in the kind of elastic wave possesses a potential energy (wave velocity, called the sonic velocity, is constant for given fluid)

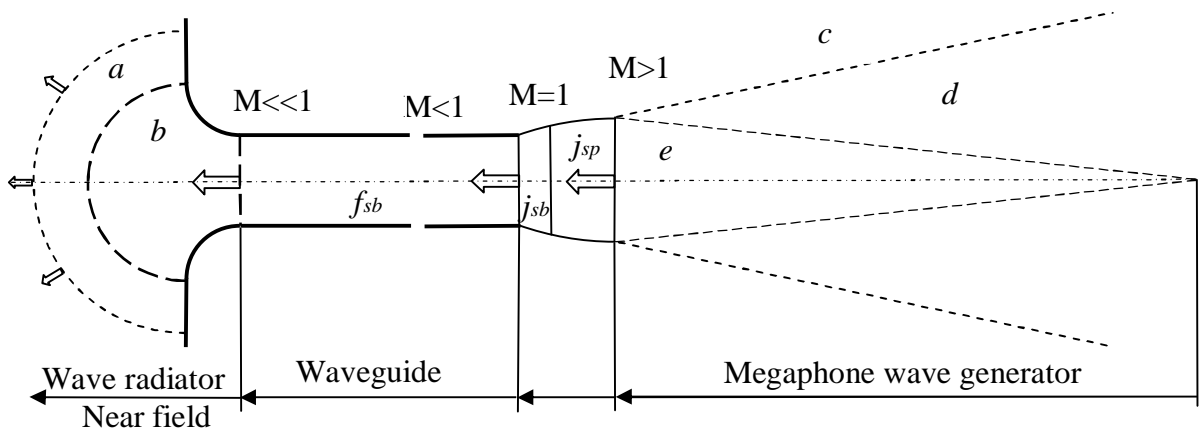

Fig.15. The gas supersonic motion: simplified scheme of the massless controlling energy transference directed against the gas stream current